\pdftrailerid{} 
\documentclass[conference]{IEEEtran} 
\IEEEoverridecommandlockouts
\usepackage{cite}
\usepackage{amsmath,amssymb,amsfonts}
\usepackage{graphicx}
\usepackage{textcomp}
\usepackage{balance}
\usepackage{xcolor}
\usepackage{url}

\def\BibTeX{{\rm B\kern-.05em{\sc i\kern-.025em b}\kern-.08em
    T\kern-.1667em\lower.7ex\hbox{E}\kern-.125emX}}

\makeatletter
\def\footnoterule{\relax%
  \kern-5pt
  \hbox to \columnwidth{\hfill\vrule width 0.5\columnwidth height 0.4pt\hfill}
  \kern4.6pt}
\makeatother

\renewcommand{\IEEEkeywordsname}{Keywords}

\begin{document}

\title{Accelerating AI Development with Cyber Arenas
  \thanks{
Research was sponsored by the Department of the Air Force Artificial Intelligence Accelerator and was accomplished under Cooperative Agreement Number FA8750-19-2-1000. The views and conclusions contained in this document are those of the authors and should not be interpreted as representing the official policies, either expressed or implied, of the Department of the Air Force or the U.S. Government. The U.S. Government is authorized to reproduce and distribute reprints for Government purposes notwithstanding any copyright notation herein.
}
}

\author{\IEEEauthorblockN{
William Cashman$^1$, Chasen Milner$^1$, Michael Houle$^2$, Michael Jones$^2$, \\ Hayden Jananthan$^2$, Jeremy Kepner$^2$, Peter Michaleas$^2$, Alex Pentland$^2$
    \\
    \IEEEauthorblockA{
    $^1$USAF, $^2$MIT
    }}}

\maketitle

\IEEEtitleabstractindextext{
\begin{abstract}
  AI development requires high fidelity testing environments to effectively transition from the laboratory to operations.  The flexibility offered by cyber arenas presents a novel opportunity to test new artificial intelligence (AI) capabilities with users.  Cyber arenas are designed to expose end-users to real-world situations and must rapidly incorporate evolving capabilities to meet their core objectives.  To explore this concept the MIT/IEEE/Amazon Graph Challenge Anonymized Network Sensor was deployed in a cyber arena during a National Guard exercise.
\end{abstract}
}

\IEEEpeerreviewmaketitle
\IEEEdisplaynontitleabstractindextext


\section{Introduction}

The increased complexity of the interactions between cyberspace and cyber-operators drive the complexity of the platforms emulating this relationship.  ``Cyber ranges'' have been used to test, develop, demonstrate, and train in cyberspace.  Expanding these simulations to cover more facets for second and third order ripple effects across domains requires more complex platforms.  These new ``cyber arena'' capabilities are better imitations of broader cyberspace in the operational environment \cite{karjalainen2020}.

Cyber arenas may be a good platform for introducing AI tools to personnel.  Adjacent roles can observe the effects of operation, including leaders, acquisitions, ethics, safety, and legal \cite{flournoy2020, ding2025}. To test the benefits of a prototype AI network sensor running in a cyber arena, the MIT/IEEE/Amazon Graph Challenge Anonymized Network Sensor was ported and run in the Persistent Cyber Training Environment (PCTE) during the Cyber Yankee annual exercise.

\section{Cyber Arenas \& PCTE}

For decades it has been recognized that modeling and simulation methods can contribute to the science of cyber security \cite{jasons2010}.  The first ``cyber ranges'' were developed in the United States by the Defense Advanced Research Projects Agency (DARPA).  Subsequent cyber ranges have since been made to demonstrate certain effects or interact with specific environments limiting their scope.  Bringing these disparate components together into a heterogeneous and modular ``cyber arena'' adds complexity, but offers more realistic emulation of cyber effects in the operational environments.  Further enhancing the experience, tangible hardware in physical space is coupled  with live or emulated users in virtual space \cite{lathrop2023, karjalainen2020}.  Significant attributes of cyber arenas include:
\begin{enumerate}
\item Realistic complexity and dependencies
\item Isolated and controlled environment
\item Internet simulation
\item User and network traffic generation
\item Attack simulation
\item Organizational personnel and infrastructure
\item Collaboration and modularity
\item Monitoring, metrics, and analysis
\end{enumerate}

PCTE is a platform developed by the Department of Defense (DOD) to provide these attributes.  For increased fidelity PCTE can connect to a derelict city with utilities. The Muscatatuck Urban Training Complex (MUTC), when connected digitally is referred to as Cybertropolis, and layers the cyber domain on the phyiscal domain to create a digital-physical multi-domain environment.  Exercises in PCTE can include interactions with the urban environment and its critical infrastructure either physically or through WiFi, Bluetooth, and cellular service.  PCTE is accessible through a web portal allowing any organization with an internet connection to participate \cite{deckard2018}.  

\section{Cyber Arenas for AI Development}

AI and Machine Learning (ML) have expanded rapidly, appearing as enhancements to other technologies or as stand-alone agents.  The DOD adoption of AI/ML has been slowed by a lack of agile Test \& Evaluation, Validation \& Verification (TEVV) environments. Cyber arenas can partially fill this AI/ML TEVV gap for the DOD.   Having a safe controlled environment can allow AI/ML to fail and be observed failing, helping prepare for failure conditions or retraining the model. Users experimenting with AI/ML tools will reveal weaknesses and build trust, especially when exposed to realistic conditions.  Metrics and analysis are part of cyber arenas so the results of every action can be scored and evaluated \cite{karjalainen2020, flournoy2020, jasons2010, ding2025}.

AI/ML tested via PCTE enables shared validations to accelerate the AI approval process.  AI/ML integration creates a ``system of systems''  enabling PCTE to act as the digital-twin for a substantial software ecosystem enabling applications to be better aligned to their potential deployed environments \cite{flournoy2020, ding2025}.

\section{Case Study: AI Network Sensing in PCTE}

The Graph Challenge Anonymizing Network Sensor is a new AI software tool to collect AI ready network data.  There are three main steps, generating or parsing network logs, building GraphBLAS traffic matrix files, and analyzing the traffic matrix files \cite{Jananthan2024}.  The traffic matrix files are highly compressed and efficient to analyze with modest computing hardware.

Cyber Yankee is a two-week exercise for New England National Guard units simulating a cyber attack on critical infrastructure.  The exercise utilizes PCTE running on $\sim$1500 Virtual Machines (VMs) with over 300 participants. Each of four blue teams have $\sim$200 VMs.  About a dozen of the machines are a mix of Windows or Linux servers, VyOS networking devices, and Modbus OT emulators. The rest are Windows 7 or Windows 10 workstations running user emulation to generate noise on the network.  All are purposely left with a few vulnerabilities to provide the red team with a starting point for the attack.  Every third year the exercise goes ``national'' by including CYBERCOM and expanding the network into a real-world water treatment facility at Cybertropolis.

Each blue team has 25-30 members to include state, industry, and foreign military partners through the National Guard State Partnership Program (SPP).  The whole red team is about the same size or a little larger than a single blue team, and over the last few years mostly composed of Marine Reserve.  The rest of the personnel are evaluators or range and exercise support.

\begin{figure}[ht]
	\center{\includegraphics[width=1.0\columnwidth]{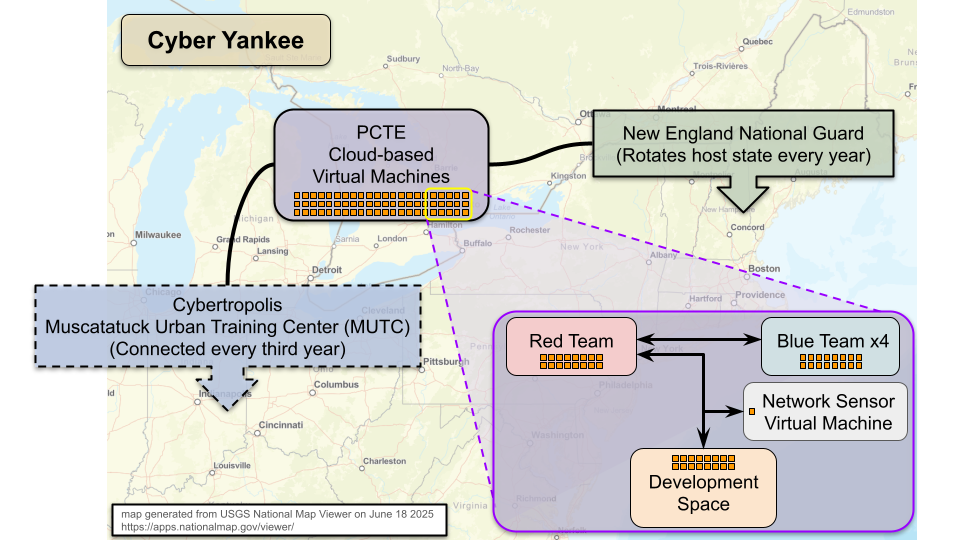}}
	\caption{{\bf Cyber Yankee Block Diagram}. The first instance of the  Anonymizing Network Sensor virtual machine (VM) was deployed in the Cyber Yankee development space.  This is where a red team would test their cyber effects and a ``Net Spec'' would be used as a template for the blue team networks.}
	\label{fig:CYBlockDiagram}
\end{figure}

The Graph Challenge Anonymizing Network Sensor was built into an Ubuntu VM.  The recommendation for the network sensor is an 8-core CPU with 32 GB RAM, when used to monitor a full  enterprise  network.  Due to the smaller scale of the Cyber Yankee network, the standard PCTE workstation allocation of a 4-core CPU and 4 GB RAM was sufficient. The VM had to be in OVA format to be uploaded.  PCTE has native support for Puppet configuration management.  For simplicity this instance of the VM used manually run Bash scripts for configuration. Tcpdump ran hourly as a service generating packet capture (PCAP) files. An hourly cron job converted the PCAP files to GraphBLAS traffic matrix files. 

The VM was run as a sidecar during the Cyber Yankee 2025 exercise (see Figure~\ref{fig:CYBlockDiagram}).  As in real networks, intermittency was encountered with port mirroring and other services.  When data was transmitted the VM's resources readily kept up with the processing load of $\sim$20 MB per hour of PCAP files that would become compressed into $\sim$6 KB of GraphBLAS traffic matrix files, yielding over 3,000x  compression.

Working with the  Cyber Yankee exercise provided valuable exposure for the Graph Challenge Network Sensor. There was realistic interference and troubleshooting.  This instance used mirrored network ports to capture network traffic, so with limited mirror ports due to network issues it created surprise AI availability events.  Although it limited data collection the event provided lessons and tuned expectations for the future.

Significant feedback from operators and observers was obtained.  An unintended utility of such compressed network matrix files could be used on low-bandwidth satellite-based networks, allowing for logging where it was previously impractical. One operator suggested a novel idea, using the network analysis technique on operating system processes and offered a new dataset for testing.  Two software representatives showed interest in the sensor as a third-party plug-in.  Multiple exercise participants in various roles expressed interest and gave comments as well when shown the sensor.

The people and experiences of Cyber Yankee provided ideas to try for next time.  The stand-alone sensor VM user-experience can be improved for collection and analysis.  Third-party plug-ins can be tried for the software already used in Cyber Yankee.  Placing sensors in the red and blue enclaves could collect more data.  If successful for multiple exercises it could be used to look for trends.

\section{Summary}

Cyber arenas expanded features better emulate cyberspace in the operational environment.  Incorporating users earlier in development should result in better software and fewer errors during real operations.  Availability will encourage experimentation and build trust in AI/ML tools.  The prototype MIT/IEEE/Amazon Graph Challenge Anonymizing Network Sensor experienced these effects by participating in an exercise on the DOD's cyber arena PCTE \cite{karjalainen2020, jasons2010, deckard2018, lathrop2023, flournoy2020, ding2025}.

\bibliographystyle{ieeetr}
\bibliography{CyberArenas}

\begin{thebibliography}{1}

\bibitem{karjalainen2020}
M.~Karjalainen and T.~Kokkonen, ``Comprehensive cyber arena; the next
  generation cyber range,'' in {\em 2020 IEEE European Symposium on Security
  and Privacy Workshops (EuroS\&PW)}, pp.~11--16, 2020.

\bibitem{flournoy2020}
M.~Flournoy, A.~Haines, and G.~Chefitz, ``Building trust through testing,''
  {\em Center for Security and Emerging Technology}, 2020.

\bibitem{ding2025}
J.~Ding, ``Machine failing: How systems acquisition and software development
  flaws contribute to military accidents,'' {\em Texas National Security
  Review}, vol.~8, no.~1, pp.~9--29, 2025.

\bibitem{jasons2010}
JASON, ``Science of cyber-security,'' in {\em Technical Report JSR-10-102}, The
  MITRE Corporation, 2010.

\bibitem{lathrop2023}
S.~D. Lathrop, ``Where is the simnet for cyberspace?,'' {\em The Journal of
  Defense Modeling and Simulation}, vol.~20, no.~3, pp.~289--294, 2023.

\bibitem{deckard2018}
G.~M. Deckard, ``Cybertropolis: breaking the paradigm of cyber-ranges and
  testbeds,'' in {\em 2018 IEEE International Symposium on Technologies for
  Homeland Security (HST)}, pp.~1--4, 2018.

\bibitem{Jananthan2024}
H.~Jananthan {\em et~al.}, ``Anonymized network sensing graph challenge,'' in
  {\em 2024 IEEE High Performance Extreme Computing Conference (HPEC)},
  p.~1–8, IEEE, SEP 2024.

\end{thebibliography}
\end{document}